# Large Exchange Magnetostriction in a Kagome Antiferromagnet at Room Temperature

Jie Du,[1,+] Liang Yao,[1,+] Hang Li,[1,*] Xiaodong Zhou,[1,2,*] Yuan Yao,[3] Xuekui Xi,[3] Yong-Chang Lau,[3] Wenhong Wang[1,*]

[1]Institute of Quantum Materials and Devices, School of Electronics and Information Engineering, Tiangong University, Tianjin 300387, China

[2]School of Physical Science and Technology, Tiangong University, Tianjin 300387, China

[3]Institute of Physics, Chinese Academy of Sciences, Beijing 100190, China

+ These authors contributed equally to this work.

* Corresponding author: hangli@tiangong.edu.cn; zhouxiaodong@tiangong.edu.cn; wenhongwang@tiangong.edu.cn

Funding: J. D. and L. Y. have contributed equally to this work. This work was supported by the National Key R&D program of China (No. 2022YFA1402600), National Natural Science Foundation of China (Grants Nos. 12204347, 12274321, 12274438, 12074415, 12361141823 and 12304066) and Beijing National Laboratory for Condensed Matter Physics (2023BNLCMPKF011). A portion of this work was carried out at the Synergetic Extreme Condition User Facility (SECUF) in Huairou Science City.

Keywords: Kagome helimagnets; exchange-driven magnetostriction; spin-lattice coupling; reversible magnetostriction

The pursuit of high-performance magnetostrictive materials is crucial for advancing technologies in sensing, actuation, and microelectromechanical systems. Although pronounced magnetostrictive effects have been observed in a few ferromagnets, a systematic exploration of magnetostriction across a broader range of antiferromagnets remains limited. Here, we report the observation of large magnetostriction in the kagome antiferromagnet $YMn_6Sn_6$. Under a magnetic field, the system undergoes a field-induced evolution from a helical magnetic ground state toward a collinear field-polarized state, accompanied by a large anisotropic lattice strain and a substantial volume magnetostriction exceeding 400 ppm at room temperature. Remarkably, the magnetostrictive response remains nearly fully reversible up to 9 T with nearly hysteresis-free behavior, effectively minimizing the energy dissipation commonly associated with domain-wall pinning. Combined experimental measurements and theoretical calculations reveal that the large, nearly hysteresis-free magnetostriction originates from the competition between intralayer ferromagnetic and interlayer exchange interactions. This exchange-driven magnetoelastic coupling further gives rise to a strongly direction-dependent lattice distortion pathway. This work establishes kagome helimagnets as a tunable platform for low-dissipation magnetoelastic functionalities and responsive magnetomechanical applications.

# 1.Introduction

Kagome magnetic materials hosting noncollinear spin textures have recently emerged as a fertile platform for exploring intertwined spin, lattice, and electronic degrees of freedom.[1–7] In these systems, competing exchange interactions and noncollinear magnetic configurations can generate strong spin-lattice coupling, enabling external magnetic fields to induce substantial lattice deformations beyond conventional spin-orbit coupling (SOC)-mediated mechanisms.[8–10] Such exchange-driven magnetoelasticity provides new opportunities for developing high-performance magnetostrictive materials with enhanced reversibility, reduced dissipation, and unconventional strain functionalities.[11–15]

Conventional magnetostrictive materials, developed from the discovery of Joule magnetostriction to modern rare-earth large magnetostrictive alloys, predominantly rely on SOC-mediated magnetic anisotropy and domain-wall motion.[16–22] Representative systems such as Terfenol-D and Galfenol exhibit important technological advantages but remain fundamentally constrained by hysteresis loss, large driving fields, mechanical brittleness, or limited strain amplitude.[14,23–27] Moreover, conventional magnetostrictive responses are often dominated by anisotropic shape deformation, while the associated volumetric lattice changes remain relatively limited, restricting their adaptability for emerging precision magnetomechanical applications. In contrast, exchange striction in noncollinear magnetic systems offers a fundamentally distinct mechanism, capable of producing volumetric lattice deformation through direct modulation of competing exchange interactions. Recent studies have revealed exceptionally large exchange-driven lattice responses in several correlated magnetic systems, including surface-enhanced strain in $Sr_4Ru_3O_{10}$,[9] spin-state-transition-induced volume magnetostriction in high-entropy alloys,[28] and large exchange striction in noncollinear antiferromagnets such as $Mn_3Sn$.[29–31] However, these effects are often accompanied by phase instability, high operating fields, strong hysteresis, or limited strain controllability.[32] Realizing large, reversible, and low-loss exchange striction in bulk materials therefore remains a central challenge.

These challenges highlight the importance of identifying bulk quantum materials that simultaneously host strong spin-lattice coupling,[33] competing exchange interactions,[34] and tunable noncollinear magnetic states.[35] In this regard, Kagome magnets have emerged as an attractive platform owing to their geometrically frustrated lattices and rich magnetic phase space. Among them, the $RMn_6X_6$ family has attracted extensive attention because of its layered Mn kagome framework and highly tunable magnetic interactions.[13,36–42] Depending on chemical composition and interlayer exchange balance, these compounds can stabilize a variety of magnetic ground states, including collinear ferromagnetic, ferrimagnetic, and noncollinear helimagnetic phases.[43–45] In particular, the coexistence of robust intralayer ferromagnetic exchange within the Mn kagome planes and competing interlayer exchange interactions along the *c*-axis provides a natural microscopic foundation for strong exchange-driven spin-lattice coupling. Such magnetic tunability makes the $RMn_6X_6$ family an ideal model system for investigating anisotropic magnetoelastic responses arising from noncollinear spin textures. Nevertheless, the microscopic mechanism linking competing exchange interactions to directional lattice deformation remains poorly understood.

Here, we report large room-temperature exchange magnetostriction in the kagome helimagnet $YMn_6Sn_6$, characterized by a large anisotropic magnetostriction approaching ~300 ppm together with pronounced volume magnetostriction of ~400 ppm at room temperature. More importantly, this magnetostrictive response remains nearly hysteresis-free up to 9 T, effectively circumventing the dissipation associated with conventional domain-wall pinning. Combining magnetostriction measurements and first-principles calculations, we demonstrate that the large lattice deformation originates from the cooperative yet highly anisotropic evolution of intralayer and interlayer exchange interactions during the field-driven spin reorientation process. Specifically, the in-plane ferromagnetic exchange favors contraction within the basal plane, whereas the evolution of interlayer exchange interactions favors elongation along the *c*-axis. By comparing calculations with and without SOC during noncollinear spin rotations, we further establish that the observed large magnetostriction is dominated by exchange interactions rather than relativistic magnetic anisotropy.

Notably, other $RMn_6Sn_6$ kagome helimagnets with strongly tilted helical spin structures, including $ScMn_6Sn_6$, $ErMn_6Sn_6$ and $TmMn_6Sn_6$, also exhibit substantial negative magnetostriction at room temperature. These findings not only uncover the microscopic origin of magnetoelasticity in kagome helimagnets but also establish noncollinear quantum magnets as a promising materials platform for low-dissipation magnetomechanical and multifunctional quantum devices.

## 2.Results and Discussion

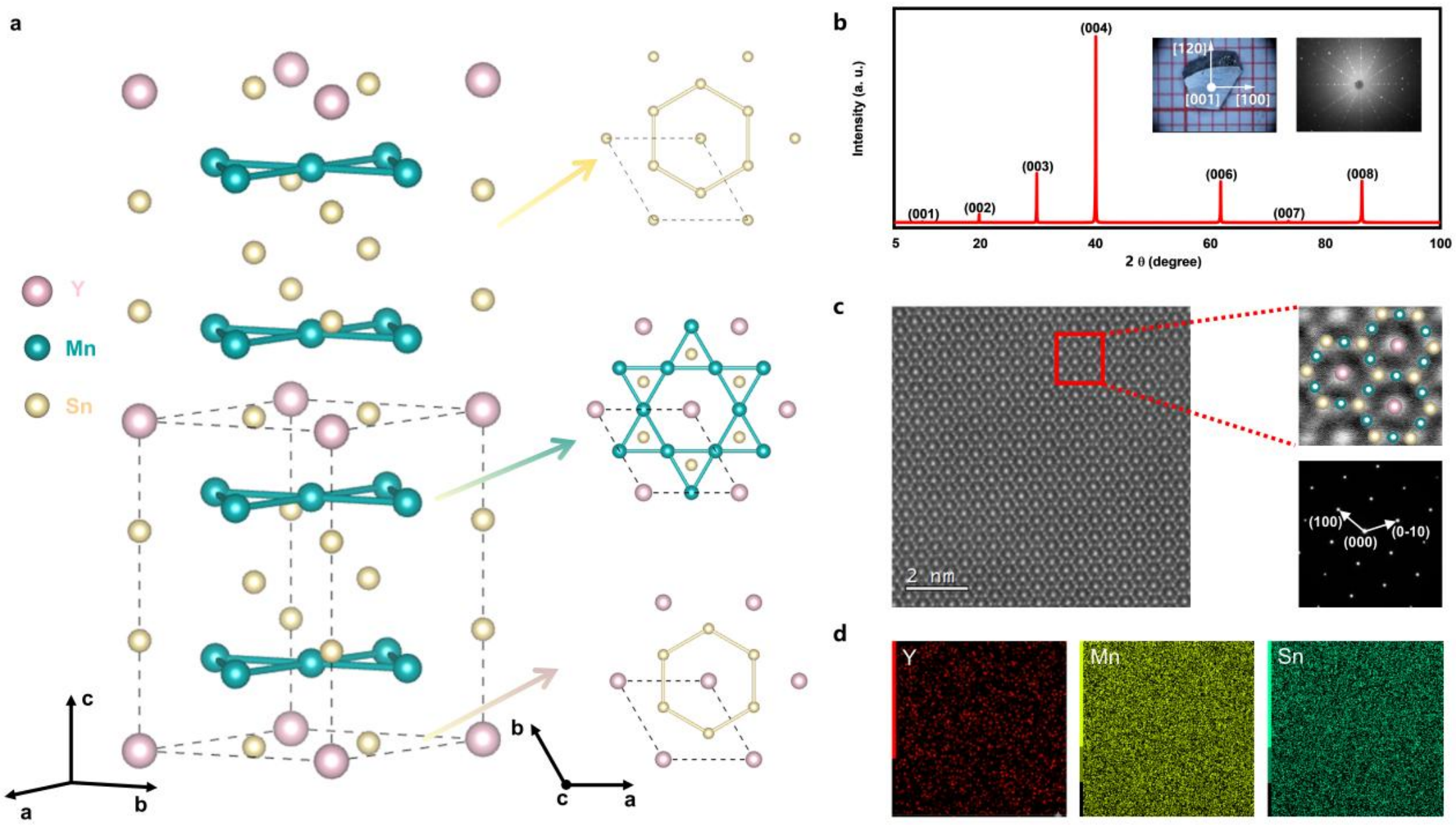

**Figure 1. Crystal Structure and crystalline quality characterization of $YMn_6Sn_6$. (a) Schematic crystal structure of $YMn_6Sn_6$. (b) Room-temperature X-ray diffraction (XRD) pattern collected from the hexagonal facet of an as-grown prismatic single crystal. The left and right insets show an optical micrograph of a representative crystal and its corresponding Laue back-reflection pattern, respectively. (c) High-resolution high-angle annular dark-field scanning transmission electron microscopy image acquired along the *ab*-plane. (d) Energy-dispersive X-ray spectroscopy elemental mapping of Y, Mn, and Sn, demonstrating high chemical homogeneity.**

**Figure** 1a presents the crystal structure of $YMn_6Sn_6$, with Y, Mn, and Sn atoms represented by pink, green, and yellow spheres, respectively. $YMn_6Sn_6$ crystallizes in the hexagonal P6/*mmm* space group (No. 191). Specifically, the Y atoms occupy the 1(a) (0, 0, 0) site, while Mn atoms reside at the 6(i) (0.5, 0, 0.25) position. The Sn atoms are distributed across three distinct Wyckoff sites: $Sn_1$ (0.333, 0.667, 0), $Sn_2$ (0.333, 0.667, 0.5), and $Sn_3$ (0, 0, 0.82). The lattice parameters are $a = b = 5.51$ Å and $c = 8.98$ Å, with $\alpha = \beta = 90°$ and $\gamma = 120°$. The layered architecture consists of YSn-Mn-Sn-Sn-Sn-Mn-YSn slabs stacked along the *c*-axis, as depicted on the right side of **Figure** 1a. Within each unit cell, Mn atoms form two well-defined Kagome layers located at z = 0.25 *c* and 0.75 *c*, providing the structural basis for competing magnetic exchange interactions and noncollinear spin textures.

**Figure** 1b displays the X-ray diffraction (XRD) and Laue back-reflection patterns of the as-grown $YMn_6Sn_6$ single crystals. The XRD pattern, collected from the hexagonal facet of the as-grown single-crystal prism, exhibits sharp peaks that can be unambiguously indexed to the (*00l*) reflections of $YMn_6Sn_6$, consistent with the crystallographic c-axis being normal to the crystal surface and confirming the high crystalline quality. The inset provides an optical image of a representative hexagonal prism-like single crystal (left) and its corresponding well-defined Laue pattern (right), which identifies the in-plane crystallographic direction as [100]. Scanning transmission electron microscopy (STEM) further corroborates these structural features. **Figure** 1c presents the high-angle annular dark-field (HAADF) images taken along the *ab*-plane, showing good agreement with the atomic structural model. Similarly, the HAADF-STEM image acquired along the *ac*-plane also agrees well with the simulated structure (Figure S1, Supporting Information). To further evaluate the chemical homogeneity and elemental distribution, energy-dispersive X-ray spectroscopy (EDS) mapping was performed. As shown in **Figure** 1d, the elemental maps for Y (red), Mn (yellow), and Sn (cyan) exhibit highly uniform spatial distributions across the examined microscale region. No discernible evidence of elemental segregation or secondary-phase precipitates was observed. The exceptional spatial coincidence of the constituent elements confirms the high crystalline quality and phase purity of the kagome samples, which is in excellent agreement with the stoichiometric integrity validated by XRD and STEM analyses.

$YMn_6Sn_6$ exhibits complex magnetic behavior characterized by multiple phase

transitions below the magnetic ordering temperature, originating from the intricate competition among distinct interlayer and intralayer Mn-Mn exchange interactions. Within the crystal lattice, while all Mn planes and intralayer nearest-neighbor Mn-Mn bonds are crystallographically equivalent due to symmetry, the interlayer Mn-Mn interactions along the *c*-axis are markedly distinct. Specifically, the Mn atoms exhibit strong ferromagnetic (FM) coupling within the kagome layers. However, the interlayer coupling is highly sensitive to the intermediate environment: it is FM when spanning the Sn-Sn-Sn triple layers, but switches to antiferromagnetic (AFM) when mediated by the Y-Sn layers.[36] These distinct interlayer exchange pathways provide the microscopic basis for competing magnetic interactions and the stabilization of the noncollinear helical magnetic state in $YMn_6Sn_6$.

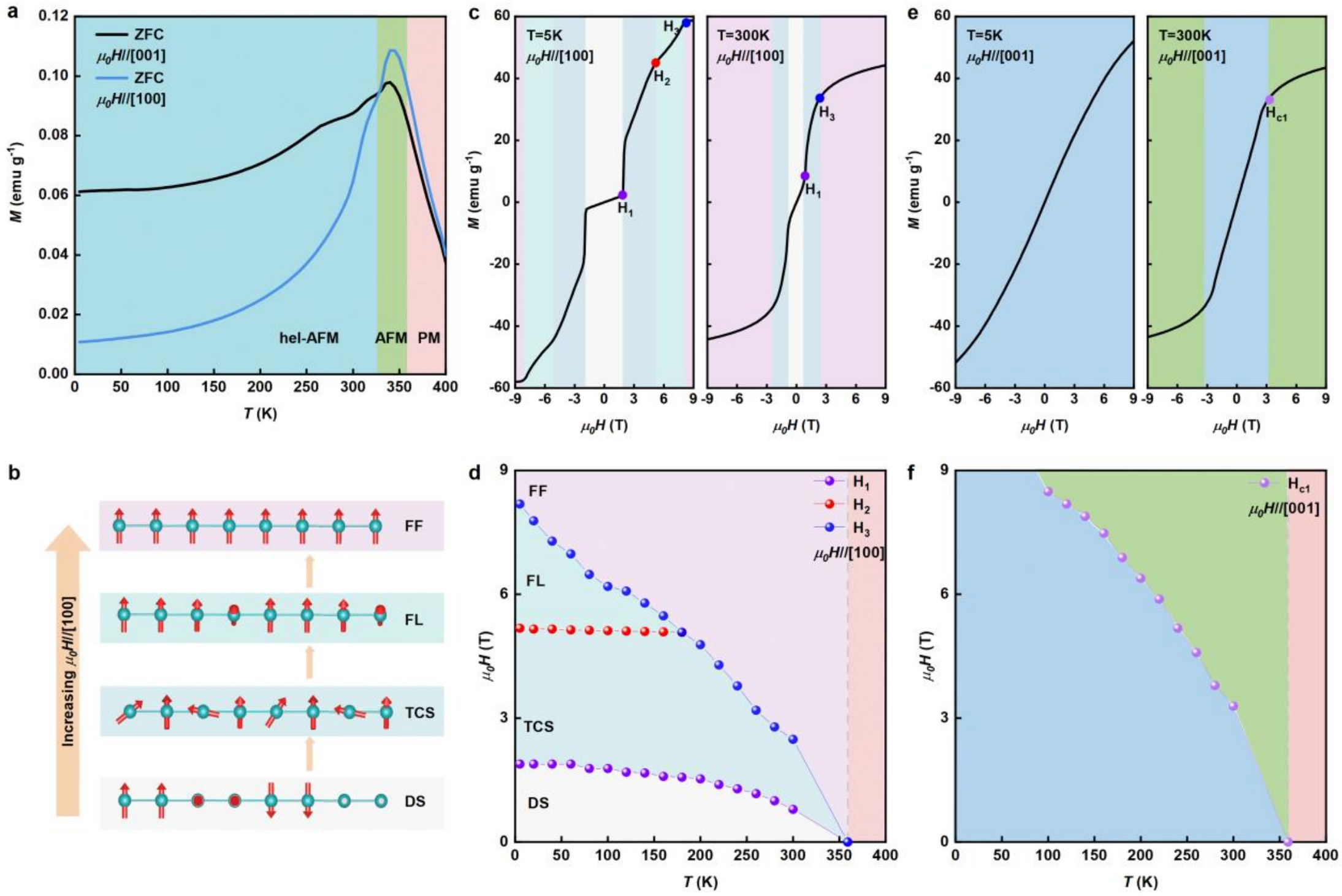

**Figure 2. Magnetic properties and phase diagrams of $YMn_6Sn_6$ under applied fields along different crystallographic directions. (a) Temperature-dependent magnetization (*M-T*) curves measured under a 0.01 T field applied along the [001] and [100] directions. (b) Schematic illustration of the field-induced rotation of Mn moments. (c) *M-H* curves measured at 5 K and 300 K with the magnetic field applied along the [100] direction. (d) Magnetic phase diagram for $\mu_0 H$ // [100], constructed by extracting the transition fields at various temperatures. (e) *M-***

***H* curves measured at 5 K and 300 K with the field along the [001] direction. (f) Magnetic phase diagram for $\mu_0 H$ // [001], constructed by extracting the transition fields at various temperatures.**

**Figure** 2a presents the temperature-dependent magnetization (*M-T*) curves under a 0.01 T field applied both in-plane and out-of-plane. Below $T_N$ = 359 K, the system initially establishes a commensurate collinear AFM structure with a propagation vector $k$ = (0, 0, 0.5). Upon cooling, an incommensurate phase rapidly emerges and briefly coexists with the commensurate phase before becoming the sole dominant phase below 300 K. Previous neutron diffraction studies indicated that this incommensurate state possesses nearly degenerate wave vectors, described as a "staggered spiral" or "double flat spiral" configuration.[36] Crucially, these noncollinear magnetic textures exhibit high anisotropic and multi-step evolution under external magnetic fields. For the in-plane configuration ($\mu_0 H$ // [100]), the magnetic moments undergo a sequential reconfiguration from a flat spiral to a transverse conical spiral (TCS), then to a commensurate fan-like (FL) state, and ultimately into the fully field-polarized (FF) state.

**Figure** 2b schematically illustrates the field-induced rotation of Mn moments when the magnetic field is applied along the [100] direction. Figure S2 (Supporting Information) shows the field-induced rotation of Mn moments for magnetic fields applied along both the [100] and [001] directions, where each disk represents one layer of Mn atoms. The *M-H* curves at 5 K and 300 K (**Figure** 2c) reveal four distinct transitions at low temperatures and three at room temperature, as highlighted by the color-coded regions. Below $H_1$, the spiral remains predominantly flat but undergoes a subtle distortion as spins tilt toward the field direction, forming the distorted spiral (DS) phase. Above $H_1$, a discontinuous jump marks the transition to the TCS phase. As the TCS cone angle diminishes at higher fields ($H_2$), the system enters a rare commensurate FL phase by flipping back to the *ab* plane to minimize the total energy. This FL phase is characterized by a quadrupled periodicity along the *c*-axis, where the spin deviation angles relative to the field direction gradually vanish until reaching the FF state. The *M-H* curves at different temperatures with the magnetic field applied along the [100] direction are shown in Figure S3 (Supporting Information). The resulting in-plane

magnetic phase diagram (**Figure** 2d) shows that the saturation field $H_3$ decreases monotonically with increasing temperature, with the FL phase disappearing above 200 K, consistent with earlier reports.[42,43] In contrast, the magnetic response for $H$ // $c$ is relatively straightforward: the helical order transforms into a longitudinal conical spiral (LCS) and eventually evolves into the FF phase (**Figure** 2e, f). The *M-H* curves at different temperatures with the magnetic field applied along the [001] direction are shown in Figure S4 (Supporting Information). Notably, while the magnetization remains unsaturated up to 9 T at 5 K due to the rigid interlayer AFM coupling, a well-defined LCS to FF transition is restored at 300 K ($H_{c1}$) owing to thermal fluctuations.

Magnetostriction is a direct manifestation of the coupling between magnetic and lattice degrees of freedom, originating fundamentally from lattice responses induced by spin structure evolution under a magnetic field. For anisotropic magnetic systems, magnetostriction is typically characterized by simultaneously measuring the longitudinal magnetostriction, parallel to the applied magnetic field, and the transverse magnetostriction, perpendicular to the field direction, thereby providing a comprehensive description of the field-induced lattice deformation. Accordingly, in the present work, strain gauges were attached along the [120] direction parallel to the magnetic field and along the [100] and [001] directions perpendicular to the field in order to systematically investigate the anisotropic magnetostrictive response of $YMn_6Sn_6$, where $\lambda_{//}$ denotes the longitudinal magnetostriction and $\lambda_{\perp}$, $\lambda'_{\perp}$ represent the transverse magnetostrictions measured along two mutually perpendicular directions.

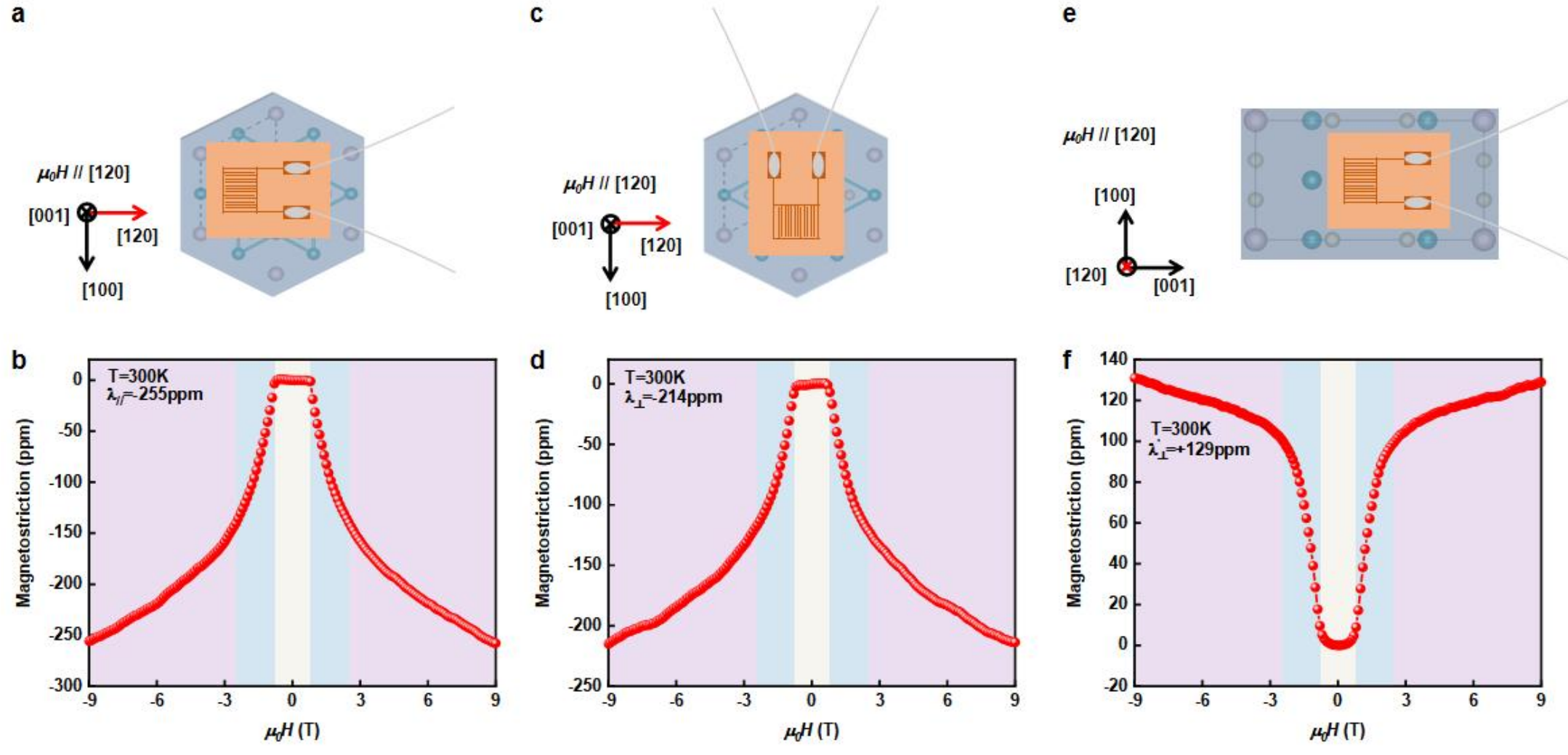

**Figure 3. Magnetostriction measurements with the magnetic field applied along the [120] direction. (a, c, e) Schematics of the measurement configurations for strain measured along the [120], [100], and [001] directions, respectively. (b, d, f) Corresponding magnetostriction data for strain along the [120], [100], and [001] directions, respectively.**

**Figure 3** presents the anisotropic magnetostriction behavior of $YMn_6Sn_6$ measured along different crystallographic directions with the magnetic fields applied parallel to the [120] direction at 300 K. Under this fixed magnetic-field configuration, where the magnetic-field orientation is maintained while the strain measurement direction is varied, pronounced and nearly symmetric field-dependent strain responses are observed for all measurement configurations, indicating strong spin-lattice coupling in this kagome helimagnet. At room temperature and within a magnetic field range of 9 T, the system delivers distinct strain responses of approximately -255 ppm along the [120] direction, -214 ppm along the [100] direction, and +129 ppm along the [001] direction. With increasing magnetic field, the strain evolves continuously and exhibits an almost hysteresis-free behavior throughout the entire field range. Such a highly reversible response suggests that the lattice deformation originates predominantly from the continuous evolution of the magnetic structure, rather than from irreversible domain-wall motion commonly observed in conventional ferromagnetic magnetostrictive materials. The longitudinal and transverse magnetostriction curves obtained with the magnetic field parallel to the [100] direction are provided in Figure S5 (Supporting Information), which exhibit excellent consistency with those measured along the [120] direction, thereby demonstrating the robustness and accuracy of the data.

Notably, the opposite signs of the lattice responses observed in the longitudinal and transverse measurement configurations reveal a pronounced anisotropy of the magnetoelastic coupling in $YMn_6Sn_6$. This anisotropy can be attributed to its layered Kagome crystal structure and the intrinsically distinct exchange interactions along the in-plane and out-of-plane directions. In $YMn_6Sn_6$, Mn atoms form ferromagnetically coupled Kagome layers within the *ab* plane, while adjacent layers along the *c*-axis are connected through relatively weak and competing interlayer exchange interactions.

Upon application of a magnetic field, the helical spin structure gradually evolves toward a more collinear magnetic state, resulting in a substantial reconstruction of the exchange energy landscape. Consequently, the competition between in-plane and interlayer exchange interactions gives rise to the experimentally observed negative in-plane magnetostriction and positive out-of-plane magnetostriction. Furthermore, the characteristic magnetic transition fields at 300 K were correlated with the magnetostriction curves and highlighted using distinct colored regions. A clear correspondence between the strain evolution and magnetic phase transition processes can be identified, further confirming that the magnetostriction in $YMn_6Sn_6$ is predominantly governed by exchange-interaction-driven spin structure reconstruction.

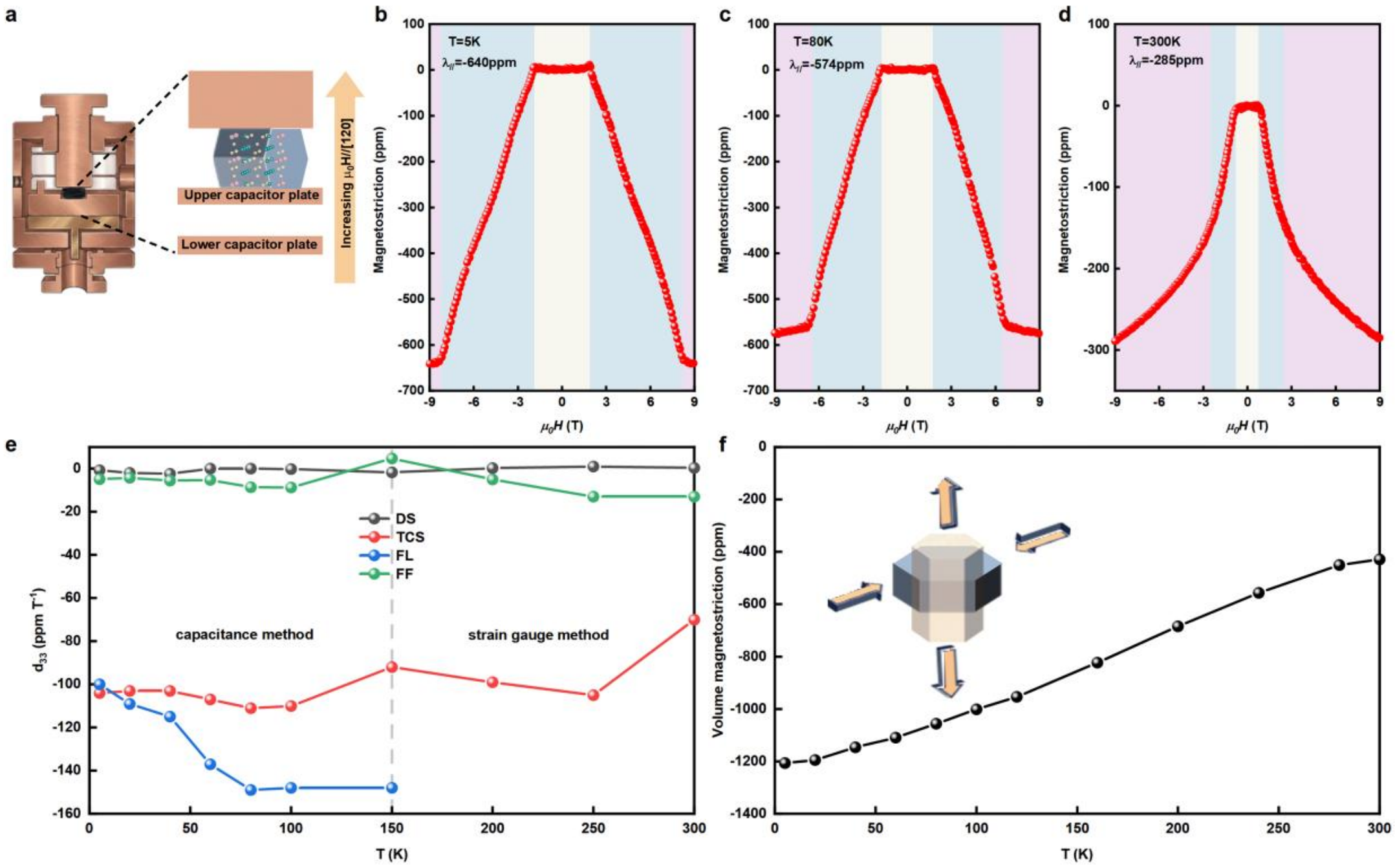

**Figure 4. Capacitance method magnetostriction measurements and temperature-dependent magnetoelastic response. (a) Schematic diagram of the capacitance method magnetostriction measurement setup. (b-d) Linear magnetostriction as a function of applied magnetic field measured at 5, 80, and 300 K, respectively. (e) Temperature dependence of the magnetostriction coefficient $d_{33}$ extracted from four different magnetic states (DS, TCS, FL and FF), comparing the results obtained by the capacitance method (below 150 K) and the strain gauge method (above 150 K). (f) Temperature-dependent volume magnetostriction w;**

**the inset schematically depicts the anisotropic lattice deformation consisting of in-plane contraction and out-of-plane expansion.**

To further verify the reliability of the magnetostriction data, independent measurements were performed using the high-precision capacitance method. The measurement principle of the capacitance method is shown in Figure S6 (Supporting Information). **Figure** 4a schematically illustrates the capacitive magnetostriction measurement setup. The upper electrode acts as a movable plate and is flexibly connected to the main frame through metallic spring sheets on the sides, while the lower electrode is rigidly fixed to the framework. The sample is mounted between the screw pillar and the sample holder, with the sample holder rigidly coupled to the upper electrode plate. When magnetostrictive deformation occurs under an external magnetic field, the dimensional change of the sample directly drives the displacement of the movable electrode, thereby modifying the gap between the two parallel plates. Since the sample deformation is directly equivalent to the variation in plate separation, ultrasensitive strain detection can be realized through capacitance measurements.

**Figures** 4b-d present the evolution of magnetostriction as a function of external magnetic field at 5 K, 80 K, and 300 K, respectively. Figure S7 (Supporting Information) shows the magnetostriction measured at various temperatures parallel to the magnetic field direction with the field applied along the [120] direction. At all temperatures, the sample exhibits characteristic nonlinear magnetoelastic behavior: the magnetostriction remains negligible in the low-field region, rapidly increases and gradually approaches a plateau at intermediate fields, and eventually reaches saturation under high magnetic fields. This behavior reflects the progressive alignment and eventual saturation of the noncollinear spin structures under field driving. A pronounced enhancement of magnetostriction is observed upon cooling. The maximum longitudinal magnetostriction extracted from the capacitance measurements increases from approximately -285 ppm at room temperature to about -640 ppm at 5 K, corresponding to nearly a twofold enhancement that even exceeds the performance of well-known Fe-Ga magnetostrictive alloy.[46] Although this value remains lower than the large magnetostriction (~1200 ppm) reported in $TbDyFe_2$ systems,[47] the present material

exhibits excellent reversibility with nearly hysteresis-free behavior, which is highly desirable for low-power magneto-actuation and high-stability functional devices. The enhanced low-temperature response suggests that suppressed thermal fluctuations facilitate more coherent spin rotations and stronger magnetoelastic coupling.

Considering that the two measurement techniques provide complementary temperature ranges for magnetostriction characterization, strain-gauge technique was employed above 150 K, whereas the capacitance method was adopted below 150 K for high-precision characterization. **Figure** 4e compares the temperature dependence of the longitudinal magnetostriction coefficient $d_{33}$ (defined as the field-induced strain per unit magnetic field along the measurement direction) obtained from both techniques. Based on the magnetic phase diagram, the magnetostriction coefficients corresponding to four distinct magnetic states were extracted. The two independent methods exhibit excellent consistency in the overall temperature evolution, both revealing a gradual weakening of magnetostriction with increasing temperature. Minor discrepancies among different magnetic states mainly originate from variations in magnetic-field sweeping paths and magnetic-state evolution, while the overall deviation remains within a reasonable range. The strong agreement between the two techniques further confirms the reliability and accuracy of the experimental results. **Figure** 4f presents the temperature dependence of the volume magnetostriction $w$, with the inset schematically illustrating the coupled in-plane contraction and out-of-plane expansion of the lattice. The volume magnetostriction is defined as:

$$w = \lambda_{//} + \lambda_{\perp} + \lambda'_{\perp}$$

The results reveal a pronounced volume contraction behavior at low temperatures, reaching approximately -1200 ppm. With increasing temperature, the magnitude of the volume magnetostriction gradually decreases but remains substantial, retaining a magnitude exceeding 400 ppm even at room temperature. This evolution is highly consistent with the linear magnetostriction results, indicating cooperative anisotropic lattice deformation on the macroscopic scale. Such pronounced and reversible volume magnetostriction, together with the nearly hysteresis-free behavior, highlights the potential of this material for low-temperature magneto-actuators, magneto-mechanical coupling devices, and highly sensitive magnetic-field sensing applications.

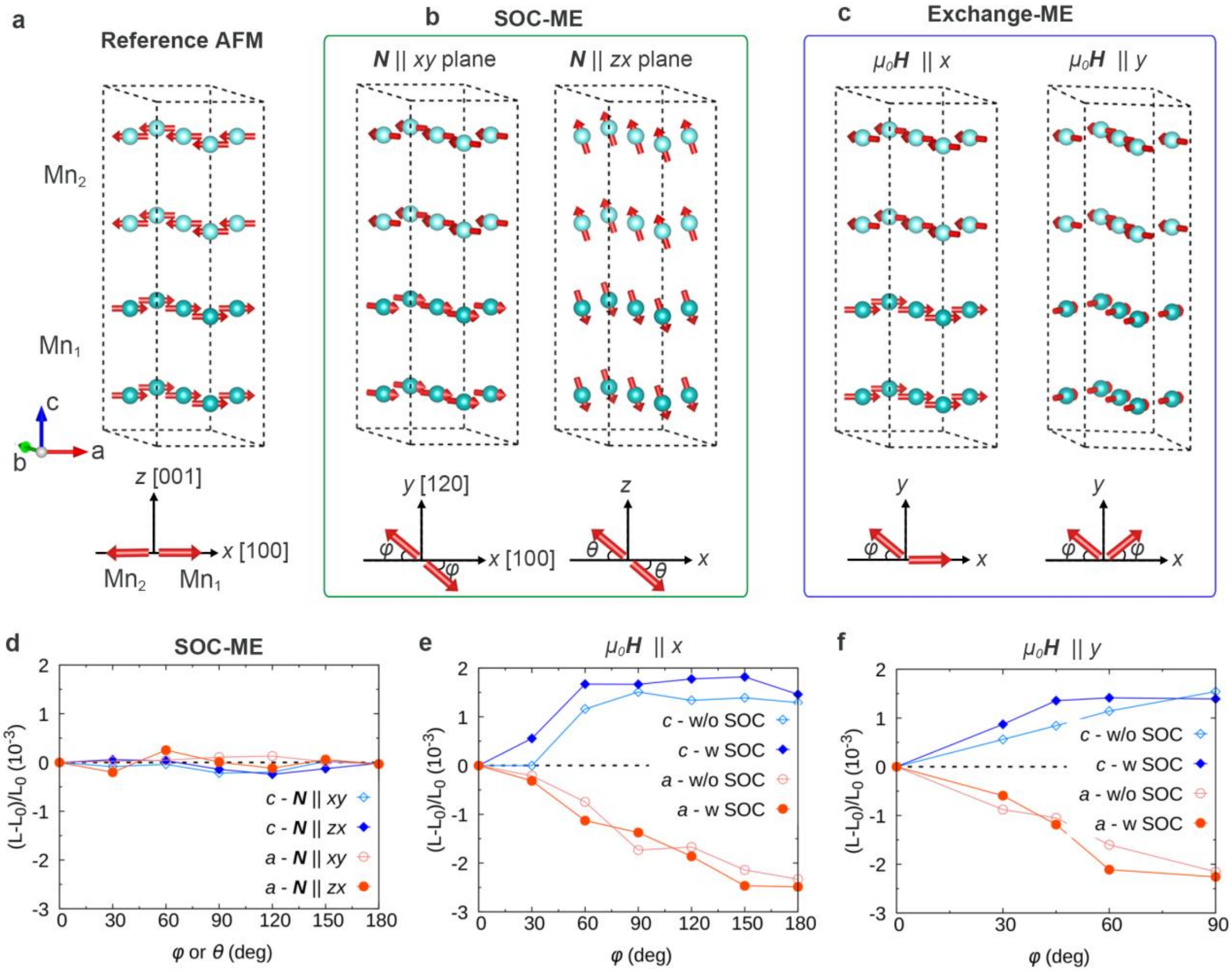

**Figure 5. First-principles calculations of exchange-driven spin-lattice coupling in $YMn_6Sn_6$. (a) Schematic of the reference AFM configuration of the minimal magnetic supercell with the Néel vector oriented along the *x* direction. (b) Simulation of the conventional spin-orbit coupling induced magnetostriction (SOC-ME). The rigid AFM spin configuration is simultaneously rotated within the *xy* plane (left, characterized by angle $\varphi$) and the *zx* plane (right, characterized by angle $\theta$) to isolate the anisotropic contributions. (c) Simulation of the exchange induced magnetostriction (Exchange-ME). Left: With the field applied along *x* axis, the $Mn_2$ sublattice gradually rotates by angle $\varphi$, driving an AFM-to-FM transition. Right: With the field applied along *y* axis, both sublattices cant towards the *y* direction by angle $\varphi$. (d) Calculated angular-dependent relative lattice strain $\triangle L/L_0$ along the *a*-axis and *c*-axis for the SOC-ME process. (e, f) Calculated $\triangle L/L_0$ with (w) and without (w/o) SOC for the Exchange-ME processes.**

To elucidate the microscopic origin of this giant anisotropic magnetostriction, we analyze the calculated lattice response following the microscopic magnetoelastic framework of Callen [48] and O'Handley.[49,50] For a given strain component $\varepsilon_\mu$, retaining the magnetoelastic coupling to first order in $\varepsilon_\mu$ and the elastic energy to quadratic order, the magnetic, magnetoelastic, and elastic contributions may be written in a single Hamiltonian as

$$H(\varepsilon_\mu) = H_{mag} + H_{ME} + H_{el} = \left[\sum_{ij} \boldsymbol{S}_i \cdot \boldsymbol{J}_{ij} \cdot \boldsymbol{S}_j + \sum_i \boldsymbol{S}_i \cdot \boldsymbol{D} \cdot \boldsymbol{S}_i\right]$$

$$+ \varepsilon_\mu \left[\sum_{ij} \boldsymbol{S}_i \cdot \frac{\partial \boldsymbol{J}_{ij}}{\partial \varepsilon_\mu} \cdot \boldsymbol{S}_j + \sum_i \boldsymbol{S}_i \cdot \frac{\partial \boldsymbol{D}}{\partial \varepsilon_\mu} \cdot \boldsymbol{S}_i\right] + \frac{1}{2} C_\mu {\varepsilon_\mu}^2$$

Here $\boldsymbol{S}_i$ is the local Mn spin at site $i$; $\boldsymbol{J}_{ij}$ and $\boldsymbol{D}$ denote the two-ion exchange-interaction tensor and the single-ion crystal-field interaction tensor, respectively; $\varepsilon_\mu$ is a strain component; and $C_\mu$ is the elastic stiffness associated with the strain component $\varepsilon_\mu$. The derivatives $\frac{\partial \boldsymbol{J}_{ij}}{\partial \varepsilon_\mu}$ and $\frac{\partial \boldsymbol{D}}{\partial \varepsilon_\mu}$ describe how the corresponding magnetic interactions respond to lattice deformation and therefore define the microscopic magnetoelastic coupling. Following Refs.[48-50], the two-ion and single-ion interactions can be partitioned into isotropic and anisotropic components. Thus, minimization of the free energy with respect to strain then gives the equilibrium magnetoelastic strain in the schematic form

$$\varepsilon_\mu = -\frac{1}{C_\mu}\left[\sum_{ij} \frac{\partial \boldsymbol{J}_{ij}^{iso}}{\partial \varepsilon_\mu} \langle \mathbf{S}_i \cdot \mathbf{S}_j \rangle + \sum_i \frac{\partial \boldsymbol{D}^{iso}}{\partial \varepsilon_\mu} \langle \boldsymbol{S}_i^2 \rangle + \sum_{ij} \frac{\partial \boldsymbol{J}_{ij}^{ani}}{\partial \varepsilon_\mu} \langle \boldsymbol{S}_i^z \boldsymbol{S}_j^z \rangle + \sum_i \frac{\partial \boldsymbol{D}^{ani}}{\partial \varepsilon_\mu} \langle (\boldsymbol{S}_i^z)^2 \rangle\right]$$

Here, the superscript $z$ denotes the corresponding spin component, while the superscripts "iso" and "ani" simply denote the isotropic and anisotropic parts of the corresponding strain derivatives. The four terms describe, in order, isotropic two-ion exchange striction, isotropic single-ion contribution, anisotropic two-ion contribution, and anisotropic single-ion contribution. The first term depends only on the relative spin correlation $\langle \mathbf{S}_i \cdot \mathbf{S}_j \rangle$ and therefore does not intrinsically require spin-orbit coupling (SOC). The second term mainly reflects changes in the magnitude of the local magnetic moment; because the calculated Mn moments vary only weakly throughout the constrained-spin rotations, its field-dependent contribution is expected to be small. The latter two terms encode the anisotropic parts of the two-ion and single-ion interactions and are predominantly SOC-related, thereby giving rise to anisotropic (Joule) magnetostriction. In particular, the anisotropic single-ion contribution is a canonical microscopic source of conventional SOC-mediated Joule magnetostriction. This decomposition therefore provides a direct framework for distinguishing exchange-driven lattice deformation from conventional SOC-mediated magnetostriction.

A direct first-principles description of the complete field evolution of $YMn_6Sn_6$ is particularly challenging because its magnetic ground state is an incommensurate helix and an applied field drives a sequence of noncollinear magnetic states. Simultaneously resolving these field-dependent spin textures and fully relaxing the lattice would require large commensurate approximants together with repeated constrained noncollinear structural optimizations. To capture the leading microscopic physics, we therefore employ a minimal 1×1×2 magnetic supercell containing four Mn Kagome layers.[51] This model is not intended to reproduce the exact incommensurate wave vector or every intermediate field-induced magnetic phase; rather, it retains the essential exchange hierarchy relevant to the magnetoelastic response. This choice is consistent with previous first-principles work on $YMn_6Sn_6$, in which the 1×1×2 supercell was used as the smallest magnetic unit cell capable of retaining the key antiferromagnetic and ferromagnetic interlayer exchange couplings.[51] In the constrained-spin model, the four Mn kagome layers are grouped into two internally collinear magnetic blocks. The relevant collective degree of freedom is therefore represented by the relative orientation between these two magnetic blocks. As illustrated in **Figure** 5a, the reference configuration of the minimal model is chosen as an AFM arrangement with the bottom two Mn layers along +*x* and the top two layers along -*x*, giving a Néel vector along *x*. The minimal four-layer model therefore captures the leading exchange striction physics associated with interlayer spin reorientation. We first evaluate the conventional SOC-induced magnetostriction (SOC-ME) by rigidly rotating the two antiparallel magnetic blocks together in two representative rotation planes, one within the basal plane (*xy*) and the other from the basal plane toward the out-of-plane (*zx*) direction (**Figure** 5b).

This protocol preserves the relative spin correlation while changing the absolute spin orientation with respect to the lattice and therefore predominantly probes SOC-dependent anisotropic contributions. Although a finite SOC-driven magnetostrictive response is obtained (**Figure** 5d), its magnitude is much smaller than that of the exchange-driven strain discussed below, indicating that conventional SOC-mediated mechanisms are not the dominant origin of the observed effect.

We next evaluate the exchange-driven magnetostriction (Exchange-ME) using two complementary in-plane spin-reorientation pathways corresponding to magnetic fields along the $x$ and $y$ directions (**Figure** 5c). For $\mu_0 H \parallel x$, the $Mn_1$ block is kept along $+x$, while the $Mn_2$ block is continuously rotated from $-x$ toward $+x$ by an angle $\varphi$, evolving from an AFM configuration toward a noncollinear and then a collinear FM state. For $\mu_0 H \parallel y$, the two initially antiparallel blocks cant symmetrically toward $+y$ by an angle $\varphi$, progressively reducing their relative angle until they become collinear along $y$. All Mn moments remain within the basal plane in both pathways, while the relative spin correlations between the two magnetic blocks are continuously modified. For every constrained spin configuration, both the atomic positions and lattice parameters are fully relaxed. Remarkably, the two pathways yield the same characteristic lattice response: a pronounced contraction of the *ab* plane accompanied by an expansion along the $c$ axis (**Figures** 5e and 5f), consistent with the experimentally observed strain anisotropy. More importantly, these large deformations persist when SOC is completely switched off, whereas restoring SOC produces only minor changes. The agreement between the $\mu_0 H \parallel x$ and $\mu_0 H \parallel y$ pathways further demonstrates that the calculated lattice response is a robust consequence of the evolving exchange spin correlations rather than an artifact of a particular rotation path. Taken together, the weak SOC-ME response, the large exchange-ME strain retained without SOC, and the consistent behavior of both exchange-driven pathways strongly support isotropic two-ion exchange striction as the dominant microscopic contribution to the magnetostriction in $YMn_6Sn_6$, with conventional SOC-mediated contributions playing a secondary role.

To assess the broader relevance of this behavior, we further performed magnetostriction measurements on several other $RMn_6Sn_6$ ($R$ = Sc, Er, Tm) helimagnets, and the corresponding experimental data are presented in Figures S8-S10 (Supporting Information). Figure S11 (Supporting Information) summarizes the magnetostriction responses of the $RMn_6Sn_6$ kagome helimagnet family measured at room temperature and at a low temperature of 5 K. Notably, compounds with different noncollinear helical spin configurations, such as $ScMn_6Sn_6$, $YMn_6Sn_6$, $ErMn_6Sn_6$, and $TmMn_6Sn_6$, all exhibit substantial negative magnetostriction with magnitudes approaching ~300 ppm at room temperature. Upon cooling, $ScMn_6Sn_6$ and $YMn_6Sn_6$ maintain negative responses, whereas $ErMn_6Sn_6$ and $TmMn_6Sn_6$ show positive magnetostriction at low temperatures. It is further observed that the sign change of the magnetostriction coincides with a change in the magnetic state at the corresponding temperature. At present, we believe that this sign reversal is intimately related to the magnetic properties, and detailed investigations are currently underway. This systematic evolution of magnetostriction, which is strongly correlated with the underlying noncollinear spin configurations, cannot be fully understood within the conventional SOC-driven magnetostriction framework. Instead, it reflects the intricate, angle-dependent competition between intralayer ferromagnetic exchange interactions within the Mn kagome planes and interlayer exchange couplings along the *c*-axis. The systematic evolution observed across the $RMn_6Sn_6$ series demonstrates that the balance between competing exchange channels and the direction-dependent lattice compliance can be continuously tuned through chemical composition, providing an effective strategy for engineering anisotropic lattice responses in noncollinear quantum magnets. These results establish kagome helimagnets as a highly tunable platform for exchange-striction-driven magnetostriction and highlight their potential for low-loss magnetoelastic functionalities beyond conventional rare-earth magnetostrictive alloys.

## 3.Conclusion

In summary, we demonstrate a large exchange-driven magnetoelastic response at room temperature in the kagome quantum helimagnets $RMn_6Sn_6$. This effect is characterized by a remarkable anisotropic strain of approximately 300 ppm and a highly reversible, near-hysteresis-free volumetric response up to 9 T. Unlike conventional magnetostrictive mechanisms governed by relativistic SOC and dissipative domain-wall motion, the macroscopic lattice deformation observed here originates from the field-induced real-space reconstruction of noncollinear spin textures. By combining experimental measurements with first-principles calculations, we establish a unified microscopic framework that directly couples magnetic reconstruction with the intrinsic mechanical anisotropy of the material, revealing a highly directional lattice compliance. Our findings position kagome helimagnets as a low-loss and highly reversible magnetoelastic platform. This work provides a predictive strategy for expanding magnetoelastic functionalities through exchange-engineered spin topologies and opens new opportunities for energy-efficient magnetomechanical devices and precision transducer technologies.

## 4.Experimental section

Sample Preparation: Single crystals of $RMn_6Sn_6$ (R = Sc, Y, Er, Tm) were prepared by Sn-flux method with a molar ratio of R : Mn : Sn = 1 : 6 : 20. High-purity R granules (99.9%), Mn granules (99.9%) and Sn granules (99.999%) were put in an alumina crucible and then sealed into a vacuum quartz tube ( $< 5\times10^{-4}$ Pa). For $ScMn_6Sn_6$, the quartz tube was heated in a box furnace to 1000°C and maintained at this temperature for 1 day, followed by cooling to 600°C at a rate of 2°C/h. Finally, the quartz tube was inverted and inserted into a centrifuge to separate the single crystals from the Sn-flux. For $YMn_6Sn_6$, the quartz tube was heated in a box furnace to 1175°C and maintained at this temperature for 1 day, followed by cooling to 600°C at a rate of 6°C/h. Finally, the quartz tube was inverted and inserted into a centrifuge to separate the single crystals from the Sn-flux. For $ErMn_6Sn_6$, the quartz tube was heated in a box furnace to 1100°C and maintained at this temperature for 1 day, followed by cooling to 600°C at a rate of 3.5°C/h. Finally, the quartz tube was inverted and inserted into a centrifuge to separate the single crystals from the Sn-flux. For $TmMn_6Sn_6$, the quartz tube was heated in a box furnace to 1000°C and maintained at this temperature for 1 day, followed by cooling to 600°C at a rate of 2°C/h. Finally, the quartz tube was inverted and inserted into a centrifuge to separate the single crystals from the Sn-flux.

Structural and Morphology Characterizations: The crystallographic orientation of the as-grown single crystals was examined using a single-crystal X-ray diffractometer (TD-3500, Dandong Mastery Technology Co., Ltd.) with Cu $K_\alpha$ radiation and further verified by Laue X-ray diffraction (TD-3000X, Dandong Mastery Technology Co., Ltd.). In addition, scanning transmission electron microscopy (JEM-ARM200F, JEOL Ltd.) combined with energy-dispersive X-ray spectroscopy (EDS) mapping was performed at room temperature to confirm the elemental distribution and structural homogeneity of the crystals.

Magnetic characterizations: Magnetic measurements were performed using a Physical Property Measurement System (PPMS-9T, Quantum Design Scientific Instruments Co., Ltd.). At room temperature, field-dependent magnetization *M-H* hysteresis loops were measured for all samples with the external magnetic field applied along the [100], [120], and [001] crystallographic directions, with the field sweeping from 9 T to -9 T and back to 9 T. The results reveal that the basal plane serves as the easy magnetization plane with negligible magnetic hysteresis. At low temperatures, the *M-H* curves along the [100] and [001] directions were systematically recorded, based on which the magnetic phase diagram was subsequently constructed.

Magnetostriction measurements: Magnetostriction measurements and cross-validation were performed using both a capacitive dilatometer and a resistance strain gauge method, with both types of measurements carried out on a PPMS. At room temperature, we measured the magnetostriction along the [100], [120], and [001] directions using strain gauges while the magnetic field was applied along the [100] direction. We also rotated the field direction to [120] and measured the magnetostriction along the [100], [120], and [001] directions. To verify the reliability of the strain-gauge measurements, the magnetostriction parallel to the applied magnetic field was independently measured using a capacitive dilatometer and compared with the strain-gauge results, showing excellent agreement. For low-temperature magnetostriction measurements, a $YMn_6Sn_6$ single crystal was selected, and the magnetostriction parallel to the applied magnetic field direction ([120]) was systematically measured over the temperature range of 150-300 K using strain gauges. At lower temperatures, the capacitive dilatometer was employed for high-precision magnetostriction measurements. Using the capacitive dilatometer, the magnetostriction along the field direction ([120]) was measured for all samples at low temperatures.

Theoretical calculations: First-principles density functional theory (DFT) calculations were performed using the Vienna Ab-initio Simulation Package (VASP).[52] The generalized gradient approximation (GGA) parameterized by the Perdew-Burke-Ernzerhof (PBE) functional was employed for the exchange-correlation potential.[53] The projector augmented-wave (PAW) method was used with a plane-wave kinetic energy cutoff of 400 eV. To construct representative noncollinear spin configurations and isolate the dominant exchange-driven magnetoelastic response, constrained-spin calculations were performed using a 1×1×2 magnetic supercell. This minimal cell is not intended to reproduce the full incommensurate helical structure, but retains the essential interlayer exchange hierarchy relevant to the present magnetoelastic analysis. Full lattice relaxations were performed at each step to track the angular-dependent evolution of lattice parameters. The calculated lattice response reproduces the experimentally observed sign and directional anisotropy of the magnetostriction. Relativistic SOC effects were systematically included to isolate and evaluate their contribution to the magnetoelastic response. A penalty functional was employed to constrain the spin directions of Mn atoms. A dense 9×9×5 Γ-centered k-point mesh was applied for this magnetic supercell. The total energy and ionic force convergence criteria were set to strict thresholds of $10^{-6}$ eV and 0.01 eV/Å, respectively.

## Supporting Information

Supporting Information is available from the Wiley Online Library or from the author.

## Acknowledgements

J. D. and L. Y. have contributed equally to this work. This work was supported by the National Key R&D program of China (No. 2022YFA1402600), National Natural Science Foundation of China (Grants Nos. 12204347, 12274321, 12274438, 12074415, 12361141823 and 12304066) and Beijing National Laboratory for Condensed Matter

Physics (2023BNLCMPKF011). A portion of this work was carried out at the Synergetic Extreme Condition User Facility (SECUF) in Huairou Science City.

## Conflict of interest

The authors declare no conflict of interest.

## Data Availability

The data that support the findings of this study are available from the corresponding author upon reasonable request.